\documentstyle[11pt,newpasp,twoside,epsf]{article}
\def\Msun{\ifmmode M_{\odot} \else $M_{\odot}$\fi}
\def\Lsun{\ifmmode L_{\odot} \else $L_{\odot}$\fi}
\def\eg{{\it e.g.,\ }}
\def\ie{{\it i.e.,\ }}
\def\etal{{et al.~}}

\def\edcomment#1{\iffalse\marginpar{\raggedright\sl#1\/}\else\relax\fi}
\reversemarginpar

\begin{document}
\title{Mapping the Dynamics of the Quasar 3C 48}
 \author{Eleni T. Chatzichristou}
\affil{NASA/Goddard Space Flight Center, Code 681, Greenbelt, MD 20771}
%\author{Ima Co-Author}
%\affil{The Name of My Institution, The Full Address of My Institution}

\begin{abstract}
The archetypical, nearby (z=0.37) quasar 3C 48 is an unusual CSS radio source
with excess far-IR emission, whose one-sided radio jet is aligned with the 
extended ionized emission and a putative second nucleus. Because of its high 
AGN luminosity and proximity, 3C 48 is a good candidate to search for kinematic
signatures of the radio jet-gas coupling and/or of a recent interaction. The 
radio morphology and our ground-based integral field spectroscopy suggest that 
the jet is interacting with its immediate environment. Using STIS aboard HST 
in several slit positions within the central 1", we map the kinematics and 
physical conditions of the extended emission line gas and their relations to 
near-nuclear star forming regions found in existing HST images. 
\end{abstract}

%\begin{figure}
%\plotfiddle{Figure1.ps}{4cm}{0}{65}{30}{-200}{-40}
%\caption{Portion of the spectrum corresponding to the central slit position, showing the $[OIII]_{4959,5007}$ emission lines. On the x (spectral) axis the wavelength increases to the right and on the y (spatial) axis N is indicated by increasing ordinates. Note the multiple line components and complex velocity field: a blueshifted component dominates $\sim$0.5 arcsec north of the nucleus while a redshifted components peaks further out and to the south.}
%\end{figure}

\section{Introduction}

In quasars and radio galaxies there is a relation between the properties of 
the radio source and those of the (optical) line-emitting gas (\eg Miley 1983; 
Tadhunter, Fosbury, \& Quinn 1989; Jackson \etal 1995).
Compact steep-spectrum (CSS) radio sources often show disturbed kinematics 
of the interstellar gas coupled to associated radio components, providing 
evidence for gas outflow. The interaction between radio jets and the host 
galaxy's ISM may be responsible for the jet confinement to small scales and 
can contribute to the ionization of the medium and lead to star formation
(\eg Axon \etal 1998; Clark \etal 1998; Villar-Martin \etal 1998 and 1999). 
Thus, detailed study of 
the circumnuclear regions is essential in understanding the processes that 
drive the kinematics, physical conditions and morphology of the line emitting 
gas.

\begin{figure}
\vbox{\plotfiddle{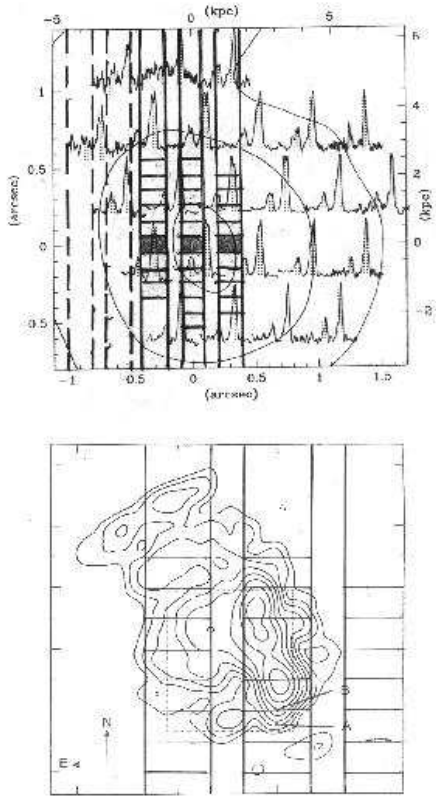}{80mm}{0}{42}{42}{-225}{-82}\vspace{-86mm}}
\hfill\parbox[b]{80mm}{\caption{ {\sl upper panel}: Spatial distribution of the $[OIII]_{5007}$ 
emission line 
profiles overplotted on a $[OIII]_{5007}$ contour map of 3C 48 (CFHT data from 
Chatzichristou
\etal, 1999). For the axes notation in kpc, we adopt $z$=0.368, H$_{0}$=75 km 
s$^{-1}$ Mpc$^{-1}$ and q$_{0}$=0. Overplotted are the (five) positions of the
STIS long slit oriented N-S (in what follows we will show data for the three 
slits drawn with full lines here). The horizontal bars within each slit 
represent 0.1{\arcsec} intervals (sum of two raw detector pixels). The dark 
areas represent the reference position of the ``nuclear'' spectrum (see also
Figure 2). {\sl lower panel}: Contours represent a VLBI map of 3C 48 
(Wilkinson \etal 1991). The ticks mark intervals of 0.2{\arcsec}. 
Overplotted are the three slit positions for which we present spectra in 
Figure 2. As before, the horizontal bars within each slit represent
0.1{\arcsec} intervals.}}
\end{figure}

\begin{figure}
\vbox{\plotfiddle{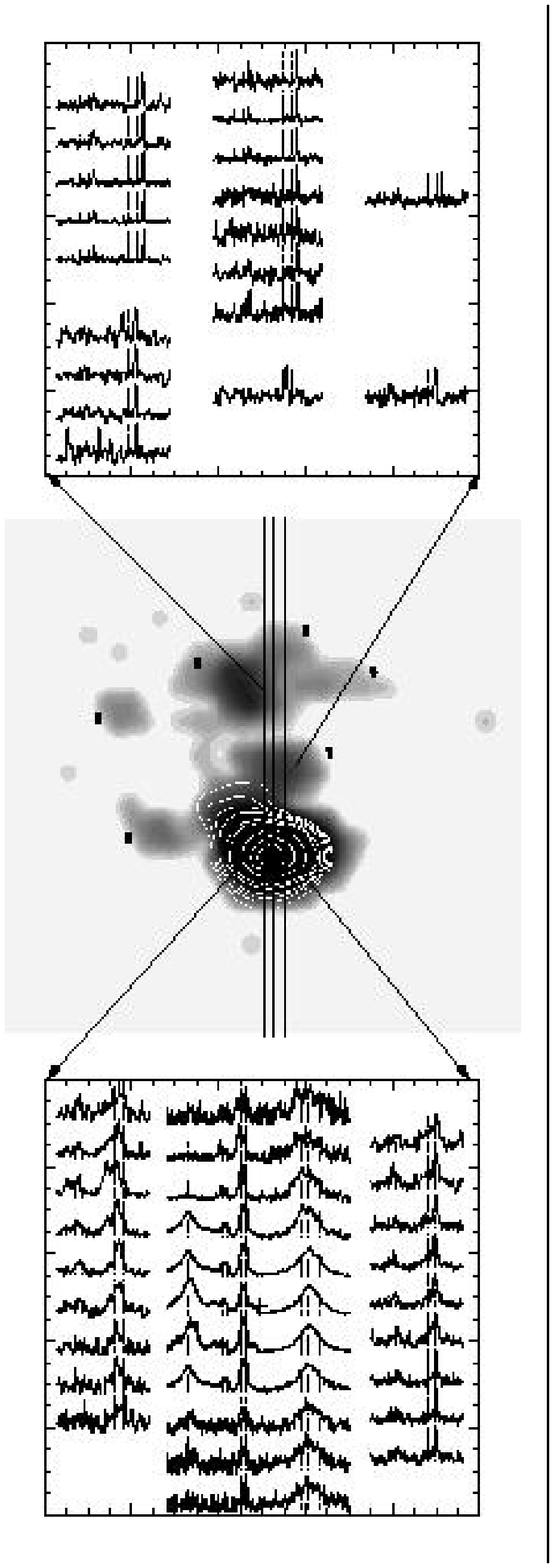}{150mm}{0}{70}{70}{-245}{-150}\vspace{-110mm}}
\hfill\parbox[b]{78mm}{\caption{The middle panel represents a (reconstructed) $[OIII]_{5007}$ 
emission image with superposed contours of a reconstructed continuum image,
of the central 6 arcsec$^{2}$ region of 3C 48 (CFHT data from Chatzichristou 
\etal 1999).
Some of the brightest emission clouds are identified and numbered. Overplotted
are three of the STIS slit positions: through the nucleus and 0.3{\arcsec} 
offset E and W. On the lower panel we show the STIS spectra in the 
central $\sim$1{\arcsec} (each spectrum corresponds to 0.1{\arcsec}). The cross
indicates the ``nuclear'' reference position (peak of emission), shown as
black areas in Figure 1. Note that for the central slit we show the $H\beta, 
[OIII]_{4959,5007}, [NII]_{6548,6583}, H\alpha$ spectral range whereas for the
two offset slits we only show the $[OIII]$ emission lines. On the 
upper panel we see the spectra corresponding roughly to the areas of the 
emission clouds 1 and 2. Again, each spectrum is summed over 0.1{\arcsec}. 
Blank areas correspond to regions of low S/N. The vertical dashed lines 
indicate the emission line positions for z=0.37 (systemic). Note that for the 
$[OIII]_{4959,5007}$ lines the position of a second, blueshifted component at 
z$\sim$0.367, is also indicated.}}
\end{figure}

\section{Previous and Current Work}

3C 48 harbours an unusually steep-core ($\alpha\approx$0.7) powerful compact 
radio source deeply embedded in the host galaxy. Its VLBI radio structure 
(Figure 1 lower panel) comprises a relatively weak core, elongated N-S and a 
powerful one-sided jet to the north, highly disrupted at $\sim$0.05{\arcsec} 
from the core, through collisions with the dense interstellar medium. This jet
then expands out to $\sim$1{\arcsec} NE \ie within the body of the host 
galaxy.
About 1.3{\arcsec} NE from the main nucleus of 3C 48, \ie roughly coinciding
with the edge of the radio jet, a high surface brightness region (3C 48A) was
detected in stellar continuum light, which was tentatively identified
with a second nucleus in the process of capture (Stockton \& Ridgway 1991;
Hook \etal 1994). In a recent paper (Chatzichristou, Vanderriest, \& Jaffe 1999) 
we have shown that this is 
more likely a region of intense star formation, triggered by the interaction 
of the radio jet with the dense interstellar medium of the host galaxy. This 
interpretation is further supported by the diffuse, knotty morphology of 
3C 48A as seen in HST images (Project ID\#05235; also, Canalizo \& Stockton 2000).

Using the technique of integral field spectroscopy at the CHFT, we have 
mapped the dynamics and emission line properties of the extended ionized gas
of 3C 48 at a resolution of $\sim$0.7{\arcsec}. In particular, we have found
that the narrow emission line profiles in the central ($\sim 1\arcsec^{2}$) 
region are split in two components,  a fast moving ($\sim$580 km s$^{-1}$ with
respect to the systemic velocity), blue-shifted, broad ($\sim$1000 km 
s$^{-1}$) component, co-spatial with the radio-jet and a spatially resolved,
narrow ($\sim$400 km s$^{-1}$) component at the systemic redshift (Figure 1 
upper panel). On the basis of these results and simple energetic arguments
we argued (Chatzichristou \etal 1999) that the blue-shifted component
might represent the imprint of the jet interaction, with the ambient gas.
Its emission line ratios indicate that the dominant ionizing mechanism are 
most probably shocks driven by the radio jet into the interstellar medium.   
These results are intriguing and need further investigation, using data of
higher quality and better spatial/spectral resolution, to establish the
detailed gas kinematics in the innermost ($\sim$1-2 kpc) region. Detection of
additional velocity components coupled to the associated radio components
(\eg Figure 1 lower panel), will provide direct observational evidence for the
AGN-driven outflow and for the motions due to bow-shocks driven by the radio 
jet into the ambient gas.

Using STIS aboard HST in long slit mode, we mapped the central 1{\arcsec} 
region in steps of 0.3{\arcsec} placing the slit along the extended line
emission (N-S) previously detected $\sim$15{\arcsec} north from the nucleus.
The spectral and spatial resolution of these new data represent an improvement
by factors $\sim$5-10 compared to the previous data (\eg Figure 1, upper panel). The analysis of these data is in progress. We study the gas kinematics primarily 
through the $[OIII]_{5007}$ 
narrow emission line and use a variety of diagnostic emission line ratios 
($[OIII]/H\beta, [OI]/H\alpha, [NII]/H\alpha$) to investigate the ionization 
structure of the emitting regions. The complexity of the line profiles and the
multiplicity of detected velocity components is obvious in Figure 2.
A third high velocity component, at -1150 km s$^{-1}$, is clearly identified
and dominates the line emission at $\sim$0.5{\arcsec} NE of the nucleus.
There is a wealth of structure in the line profiles, obviously coupled to the
radio components seen in Figure 1 and on more recent VLBI data.
The full analysis and results will be presented in a forthcoming paper, where
the main questions addressed will be: what is(are) the acceleration 
mechanism(s) and where exactly in the flow is the gas accelerated? What is the
dominant ionization mechanism? Furthermore, comparison with the ionized gas 
that lies further out from the radio axis allows us to estimate the cooling 
times and thus the shock velocities involved. Comparison of these results with
existing dynamical models for the formation of NLR in AGNs will help to 
further constrain these models for larger jet powers and to clarify the regime
of jet confinement.

%\begin{figure}
%\plotfiddle{Figure3.eps}{40mm}{0}{35}{35}{-110}{+20}
%\caption{Close-up of the spectra corresponding to the central 0.75 {\arcsec}
% from
%the two offset slit positions: 0.3{\arcsec} E (left panels) and 0.3{\arcsec }W
%(right panels). From top to bottom we show spectra summed over 0.25{\arcsec} 
%(5 pixels) north, around, or south of the reference position (Figures 2 and
%3). The lowest panels indicate the sum of the above spectra, i.e. sum over 
%0.75{\arcsec}. Note the very different line profiles corresponding to the two 
%slit positions. Narrow line profiles corresponding to the systemic velocity
%on the E, while at least two more blue-shifted components appear W of the 
%nucleus.} 
%\end{figure}

%\pagebreak
%\section{Examples}

%These instructions give an overview of the basic markup commands
%that need to be entered in a paper.
%Authors are encouraged to examine
%the sample papers that are included with the style file;
%these examples are named \verb"psample1.tex"
%and \verb"psample2.tex".  The file \verb"psample1.tex" is a paper prepared
%with the PASP macros utilizing a \emphasize{minimal} amount of markup.
%A more ``complete'' paper requiring most of the capabilities
%of the package is provided as \verb"psample2.tex";
%this file is annotated with comments that describe
%the purpose of the markup.

\end{document}